# Bright squeezed light in the kilohertz frequency band

Ruixin Li[1], Bingnan An[1], Nanjing Jiao[1], Junyang Liu[1], Lirong Chen[1,2], Yajun Wang[1,2*], and Yaohui Zheng[1,2**]

[1]*State Key Laboratory of Quantum Optics Technologies and Devices, Institute of Opto-Electronics, Shanxi University, Taiyuan 030006, China*

[2]*Collaborative Innovation Center of Extreme Optics, Shanxi University, Taiyuan, Shanxi 030006, China*

[*]*e-mail: YJWangsxu@sxu.edu.cn*

[**]*e-mail: yhzheng@sxu.edu.cn*

## Abstract

The largely technical noise of a free running laser is the fundamental limit for preparation of a bright squeezed light, especially within MHz band. We construct a universal and complete theoretical model for nonclassical stabilization, and propose a novel bright squeezed light generation scheme with assistance of a hybrid noise stabilization technology. The scheme assimilates a broadband passive noise stabilization to successfully extend the feedback frequency bandwidth of a nonclassical active stabilization to MHz band. Meanwhile, the technical noise suppression magnitude is further improved by 9 dB, which creates the prerequisite conditions for preparing a broadband bright squeezed light. Finally, a -5.5 dB bright squeezed light with a power of 1 mW and a squeezing bandwidth among kHz to MHz band was generated. The experimental results agree well with the theoretical one. To the best of our knowledge, it is the first demonstration of a milliwatt-order bright squeezed light across the kHz frequency band. The demonstration sheds light on potential applications of bright squeezed light in quantum metrology and might enable new concepts in the future.

## Introduction

Squeezed states of light describes a unique quantum state where the uncertainty in a certain quadrature, amplitude or phase, is reduced below the shot noise limit (SNL) at the expense of enhancing the orthogonal quadrature[1,2]. Exploiting the sub-shot noise property, squeezed state of light has become an extremely valuable resource for various applications in quantum information technologies[3-8]. In the first case, the laser power is limited to relative low level owing to the damage threshold of sample under test, such as biological imaging[9-12] and medical diagnostic[13,14], the influence of nonlinear effect in fiber, such as fiber communication[15,16] and clock synchronization[17,18], as well as the maximum payload of satellite and airborne platform[19]. In the second case, a higher laser power inevitably induces excess technical noise, which degrades, rather than improves, the sensitivity of the system, like plasmonic sensing[20-23]. In the third case, limited by the performance of current technology, laser power cannot be arbitrarily increased, e.g., laser doppler anemometry[24].

However, except for noise suppression[25-27], the frequency band and output power of the squeezed light are also the key issues that need to be addressed to satisfy concrete application scenarios, such as biological tracking among kHz to MHz band[28-31], microcantilever displacement measurement within hundreds of kHz band[32,33], optomechanical coupling in kHz frequency range[34,35]. Moreover, all these applications mentioned above require that laser has an output power of from microwatt-to-milliwatt order.

Until now, the maximum squeezing strength is -15 dB at the MHz sideband frequencies of 1064 nm wavelength[27]. Besides the squeezing strength, extending the squeezing bandwidth to lower frequency has also gained wide interests, especially in quantum precision measurement[36] and sensing[37,38]. Currently, a key mechanism for generating squeezed light below the MHz range exploits the immunity to nonlinear noise coupling between the pump and seed beams[39-41], while preventing the transfer of the seed laser's high technical noise at low frequencies[42]. To avoid the noise coupling and transfer at low frequency band, vacuum field (no seed input) instead of a seed beam at the carrier frequency usually serves as the seed field of an optical parametric oscillator (OPO). Therefore, the OPO scheme is only suitable for the generation of squeezed vacuum states[40,41]. Unfortunately, squeezed vacuum state has rather low power, which is not enough to meet the requirements of biological and optomechanical sensing[28-35].

Bright squeezed light, defined as light exhibiting sub-shot-noise level (similar to squeezed vacuum states) combined with significant power, is crucial for improving sensitivity. By exploiting a radiation-pressure-driven interaction of a coherent light with a mechanical oscillator, a narrow-band bright squeezed light with a squeezing strength of -0.7 dB was experimentally generated[43]. The passive interference of squeezed vacuum with a laser beam on a beam splitter (BS) is a conventional scheme for preparing a bright squeezed light[11,35]. This requires a quantum-noise-limited laser beam. Otherwise, the significant low-frequency classical noise inherent in most lasers would spoil the squeezing performance. Even if the laser beam is quantum-noise-limited, a quantum noise is still transferred onto the light, due to the vacuum noise injection from the empty port of the beamsplitter and shot noise of the output laser beam. But, limited by the vacuum noise introduced by the beamsplitter, the optimum squeezing strength is constricted to a lower level. To date, only -2.6 dB of bright squeezed light with output power of 25 μW was experimentally demonstrated in the 2-200 kHz range[35]. Theoretically, active feedback with infinite gain can thoroughly compensate the noise degradation originating from the technical noise of the laser beam, i.e., extending the noise squeezing bandwidth to lower frequency, and driving the final noise limit to electronic noise of the in-loop and quantum noise penalty of the out-of-loop[44,45.] There is also a tradeoff between the lower frequency bound and the out-of-loop output power. This tradeoff can be tuned by adjusting the splitter ratio in the active feedback control loop. In general, control loop with large feedback gain is the basis of increasing out-of-loop power. Based on a pioneering theory[44], nonclassical feedback stabilization, extracting the error signal via a 50:50 BS, produced -5.7 dB squeezed light at 9.9 mW within the 6–20 kHz frequency band[46]. However, there has a contradiction between the gain and bandwidth of the control loop[47,48], limiting the squeezing bandwidth to tens of kHz order of magnitude.

In this article, we construct a universal and complete theoretical model of bright squeezed light generation, and propose a new conceptual framework to prepare a bright squeezed light among kHz to MHz band by employing a hybrid passive and active nonclassical stabilization scheme. The hybrid stabilization scheme incorporates the advantages of high loop gain and broadband noise suppression properties of the two stabilization techniques, respectively. It suppresses the technical noise of original laser beam from -125 dB/Hz to -166 dB/Hz across the kHz band, which is 10 dB lower than the relative shot noise level of -156 dB/Hz. The ultra-low intrinsic noise relaxes the control loop gain requirements, enabling extension of the loop bandwidth to the MHz range. By carrying out the nonclassical stabilization on a 99:1 BS, we generate a -5.5 dB bright amplitude squeezed light with 1 mW optical power and kHz to MHz frequency bandwidth. The experimental results agree well with the theoretical analysis. To the best of our knowledge, it is the first demonstration of a milliwatt squeezing regime in such a broadband frequency range. The demonstration sheds light on potential applications of bright squeezed light in quantum metrology and might enable new concepts in the future.

# Results

## Principle analysis

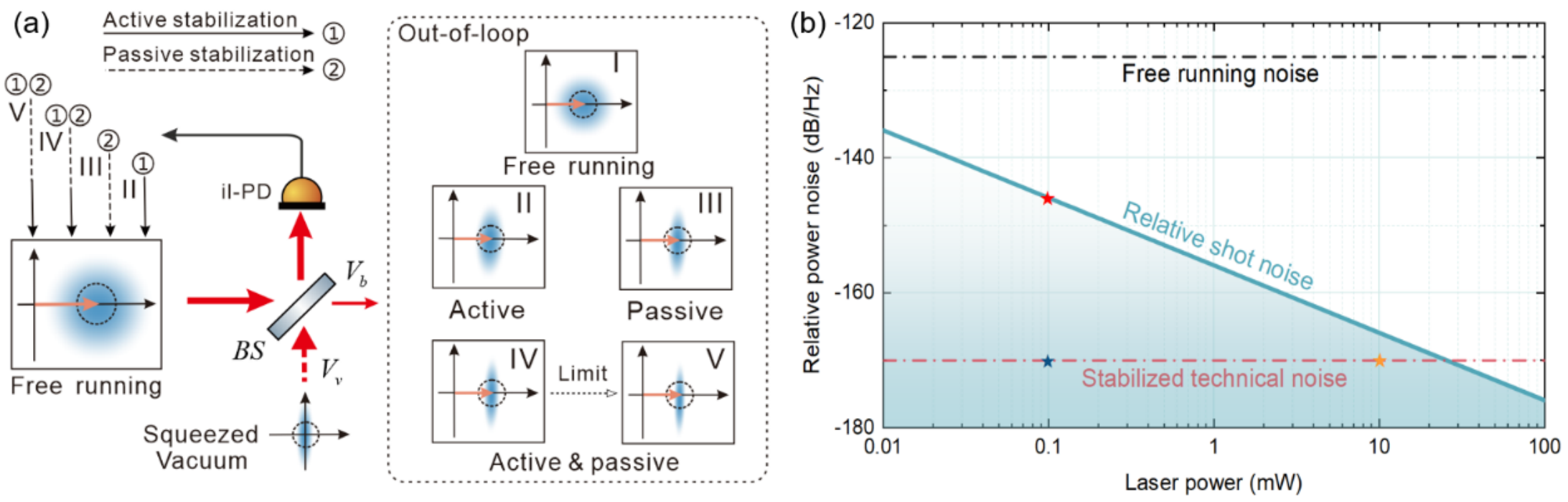

FIG. 1 Strategy for bright squeezed light generation and the technical noise compared with SNL in the system. (a) Simplified schematic of squeezed light generation with multiple noise suppression techniques. (I) Free running noise; (II) Active stabilization at detectable power; (III) Passive stabilization; (IV) Active and passive stabilization at detectable power; (V) Active and passive stabilization at the available limiting power of the system; il-PD, In-loop photodetector; BS, Beam splitter; (b) Relationship between the laser power and relative shot noise. Red star: quantum noise penalty of the out-of-loop beam; Blue star: technical noise after active feedback; Yellow star: in-loop electronic noise.

The schematic diagram of bright squeezed light generation is shown in Fig. 1(a), which is a nonclassical active feedback on a BS with assistance of a squeezed vacuum. Before the active feedback, the free-running laser is pre-stabilized by employing a passive noise stabilization technology. One of the output lasers serves as an in-loop sensing beam, which is detected by the in-loop photodetector (il-PD) to produce an error signal of active stabilization. This signal is used to suppress the amplitude noise to far below the shot noise of the out-of-loop beam. Therefore, we can extract a sub-shot noise light at the out-of-loop port. In this process, the squeezed vacuum not only suppresses the vacuum in the in-loop beam, but also provides enhanced quantum noise reduction for the out-of-loop beam, compared to the passive interference regime. Here, two types of critical noise sources are relevant to the squeezing strength. One is the technical noise in kHz frequency band of the free running laser. The other one is the quantum noise, which is imprinted by the vacuum noise originating from the dark port of the BS, and the shot noise of the light transmitted from the BS. Based on the theoretical derivation of nonclassical feedback (see Supplementary Note 2), the amplitude quadrature variance of the light transmitted from the BS can be approximately simplified to[44]

$$V_b(P_{OOL}) = \frac{V_\nu}{r} + \frac{TN_{OOL}}{RSN_{OOL}}, \tag{1}$$

$P_{OOL}$, $TN_{OOL}$ and $RSN_{OOL}$ are the power, relative technical noise (*TN*) and relative shot noise (*RSN*) of the out-of-loop laser beam (bright squeezed light), respectively, where $RSN = 2h\nu/P$ ($\nu$ is the laser frequency, and $P$ is the total input power) as shown in Fig. 1(b). $V_\nu$ is the noise variance of the squeezed vacuum, and $r$ denotes the power reflectivity of the BS. The first term of Eq. (1) represents the lower bound of the noise variance, which depends on the squeezing strength of the quantum state, as well as the splitting ratio of the BS. The second one originates from the technical noise of the laser, which should be suppressed as much as possible to far below the SNL of the out-of-loop laser beam.

In our scheme, the laser technical noise is passively pre-stabilized to near SNL among kHz to MHz frequency band by exploiting a second harmonic generator (SHG), which not only relaxes the requirements for high gain broadband feedback control loop, and also extends the squeezing bandwidth to MHz range. The noise suppression mechanism is a nonlinear three-step-photon-recycling (TSPR) process (see details in Materials and methods), generating a laser power of more than 100 mW for bright squeezed light generation[49]. After passive stabilization, the technical noise of the initial laser can be approximately expressed as

$$TN_{OOL-P} = g(f) \cdot TN_F, \tag{2}$$

where $g(f)$ is the noise reduction factor of TSPR, and $TN_F$ is the technical noise of free-running laser (see Supplementary Note 1). $g(f)$ is approximatively frequency-independent across the kHz band[49]. Subsequently, integrating with nonclassical active feedback (where $G(f) \gg 1$), the technical noise of the out-of-loop laser beam can be inferred as (see Supplementary Note 2)

$$TN_{OOL-P\&A} = g(f) \cdot TN_F \cdot \frac{1}{\left|1-\sqrt{r}\cdot G(f)\right|^2} + RSN_{IL}V_e, \tag{3}$$

where, $RSN_{IL}$ and $V_e$ are the *RSN* relating to the detected laser power and electronic noise of the il-PD, respectively. $G(f)$ is the feedback gain of active control loop, which is frequency-dependent due to an inherent time delay induced limited feedback bandwidth $f_B$. As a rule of thumb, its inherent mechanism restricts the best noise suppression to a factor of $\sim f_B/f$ at Fourier frequency $f$[47,48]. The first term of Eq. (3) denotes the residual technical noise of the out-of-loop unit. An infinite gain $G(f)$ can drive the residual technical noise to infinitesimal. The second term is the electronic noise of the feedback control loop, which mainly refers to the electronic noise of the photodiode imprinted on the out-of-loop beam. In principle, with large feedback gain, the out-of-loop technical noise $TN_{OOL}$ can be suppressed far below SNL of the out-of-loop beam. Then, the electronic noise of the photodiode ($RSN_{IL}V_e$) becomes the key limit factor to the squeezing strength, especially in the case of higher power output. It may lead the $RSN_{OOL}$ to approach the electronic noise level, and no squeezing can be observed. It follows the fact that, the electronic noise associated with the in-loop photodetector (il-PD) are imprinted onto the beam transmitted from the BS, forming the electronic technical noise limit of the out-of-loop unit.

Interestingly, if we create a condition of $r > V_\nu$ and $TN_{OOL} << RSN_{OOL}$, the scheme with assistance of a squeezed vacuum can be used to produce a sub-shot noise light, i.e., transform the squeezed vacuum into amplitude squeezed light. The key point is to suppress the $TN_{OOL}$ far below $RSN_{OOL}$. Here, we take the experimental parameters in our scheme as an example, in which the BS power reflectivity $r$ equals to 99%. A passively stabilized laser beam with a power of 10 mW is divided into two beams by the BS. The 9.9 mW laser serves as the in-loop sensing beam readout by the il-PD, and the electronic noise of il-PD is shown as the yellow star in Fig. 1(b). The optical power of the out-of-loop laser is 0.1 mW, whose quantum noise penalty ($V_v/r$) is indicated by the red star in Fig. 1(b). Under an infinite feedback gain, $TN_{OOL}$ can be completely suppressed to the electronic noise level of the il-PD, shown as the blue star in Fig. 1(b), which is 24 dB lower than $RSN_{OOL}$. Consequently, the influence of the second term of the Eq. (1) on squeezing can be neglected. Therefore, it builds the physical conditions for bright squeezed light preparation.

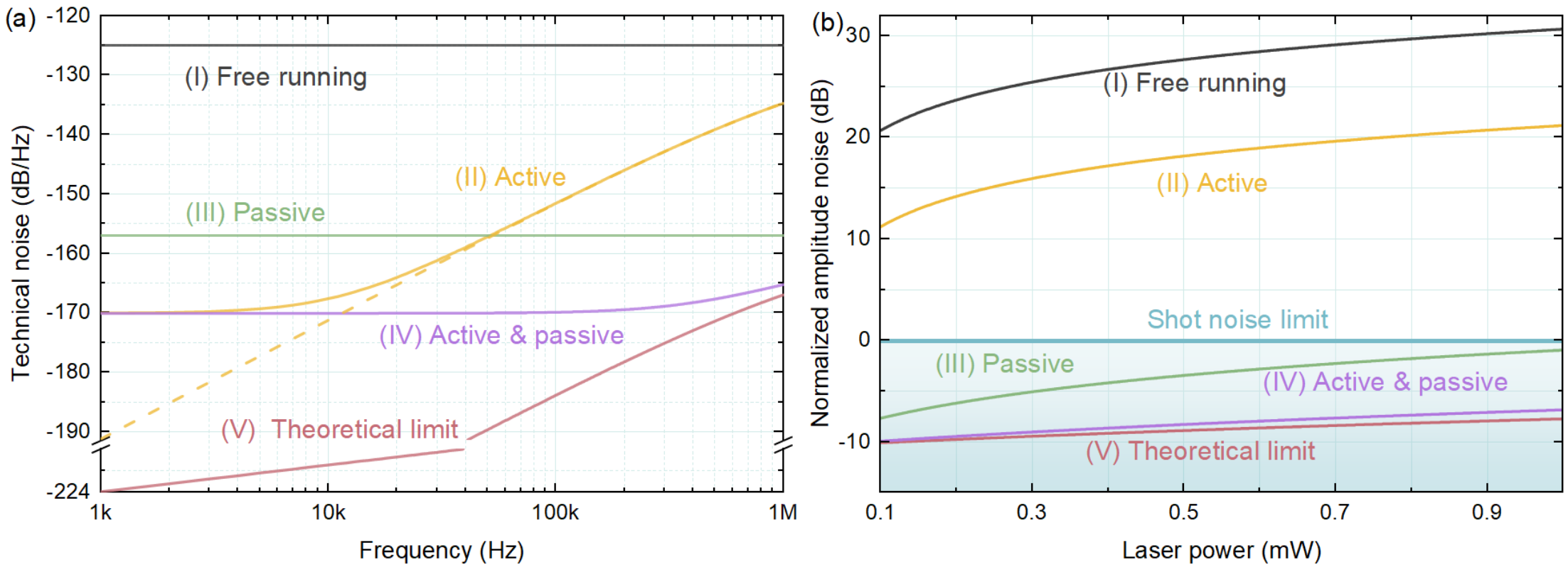

FIG. 2 Analysis results of the technical and the squeezed noises for different conditions. (a) Dependence of the out-of-loop technical noise $TN_{OOL}$ on analysis frequency. Experimental parameters are listed below: BS splitting ratio: 99:1; feedback bandwidth: 2 MHz; free running noise: -125 dB/Hz; detectable power of the in-loop: 10 mW; electronic noise of the PD is -170 dB/Hz; squeezing strength: -10.5 dB. We simulate the theoretical limit assuming full in-loop power of 99 mW and neglecting the PD's electronic noise (trace V). However, in practice, the detectable power is limited to 10 mW due to the saturation power of the il-PD, and the contribution of electronic noise becomes non-negligible (traces I-IV). The dashed line is calculated with a factor of $f_B/f$. (b) Dependence of the squeezing strength of bright squeezed light on the output power at the analysis frequency of 1 MHz, which represent the maximum measurement frequency in squeezing bandwidth. All these results are inferred by Fig. 2 (a) and Eqs. (1)-(3), which are normalized to SNL.

To elaborate the mechanism more clearly, we simulate the technical and squeezed noises for different technology based on Eqs. (1)-(3), under the experimental parameters listed in the caption of Fig. (2). Fig. 2(a) shows the theoretical results of $TN_{OOL}$. Trace (I) is the technical noise of the free-running laser at the il-PD's saturation power of 10 mW, which corresponds to an electronic noise level of -170 dB/Hz. Firstly, with a single passive noise stabilization[49], the technical noise is reduced to -157 dB/Hz (trace (III)) among kHz to MHz frequency band, with a noise suppression of 32 dB. The gap between the technical noise and electronic noise is only 13 dB, which indicates $TN_{OOL}$ is still the main limitation for bright squeezing preparation with passive interference regime. With power increasing, the squeezing strength is decreased rapidly (trace (III) in Fig. 2(b)). Secondly, by exploiting the nonclassical active noise stabilization independently, the technical noise is calculated as trace (II). The feedback bandwidth of our control loop is $f_B$ =2 MHz. The best noise suppression with optimized feedback loop for different $f$ is calculated as the dashed line of trace

(II) in Fig. 2(a), in which the feedback gain is linearly reduced along with the Fourier frequency $f$ increasing, as well as the technical noise suppression level. In real applications, the active feedback noise level is also limited by the electronic noise of the photodiode, which constrains the technical noise to the electronic noise floor with finite feedback gain below 20 kHz, shown as the solid line of trace (II) in Fig. 2(a). As $f$ is increased to 50 kHz, the technical noise is raised up to the shot noise limit level of the out-of-loop laser field, above which the squeezing is rapidly submerged in the technical noise floor, e.g., trace (II) in Fig. 2(b). Therefore, the squeezing bandwidth is always limited to within 100 kHz by using pure active feedback control regime[35,46]. Finally, we integrate the passive stabilization in the nonclassical active one, the noise suppression frequency range can be extended to MHz frequency, shown as trace (IV) in Fig. 2(a). In this case, a broadband high power bright squeezed light can be prepared within MHz band, indicated by trace (IV) in Fig. 2(a) and (b). Assuming no electronic noise limit and a passively pre-stabilized laser beam with an available power of 99 mW detected by the il-PD, $TN_{OOL}$ can be suppressed to -224 dB/Hz at 1 kHz, shown as trace (V) in Fig. 2(a).

According to Eqs. (1)-(3) and results of Fig. 2(a), we simulate the dependence of the squeezing strengths on the output power at the Fourier frequency of 1 MHz (shown in Fig. 2(b)), which represent the worst noise reduction among kHz to MHz frequency band. The results indicate that the individual noise stabilization regime is infeasible to generate a broadband bright squeezed light with MHz bandwidth and milliwatt-order output power. With nonclassical hybrid noise stabilization scheme, we expect to generate a -6.5 dB bright squeezed light among kHz to MHz frequency band with 1 mW output power. By comparing traces (IV) with theoretical limit traces (V) in Fig. 2(a) and (b), we can expect to enhance the bright squeezed light to higher squeezing strength and power simultaneously, with improved saturation power and reduced electronic noise of the il-PD. Apparently, the new perspective rooting from the theorical models allows us to comprehensively and expediently optimize the practical experimental parameters or conditions for nonclassical power noise stabilization. With this complete model, we have found that the final noise performance of the out-of-loop beam is co-determined by the residual technical noise relating to the optical field and electronic noise of the il-PD, as well as the quantum noise penalty of the out-of-loop.

## Experimental setup

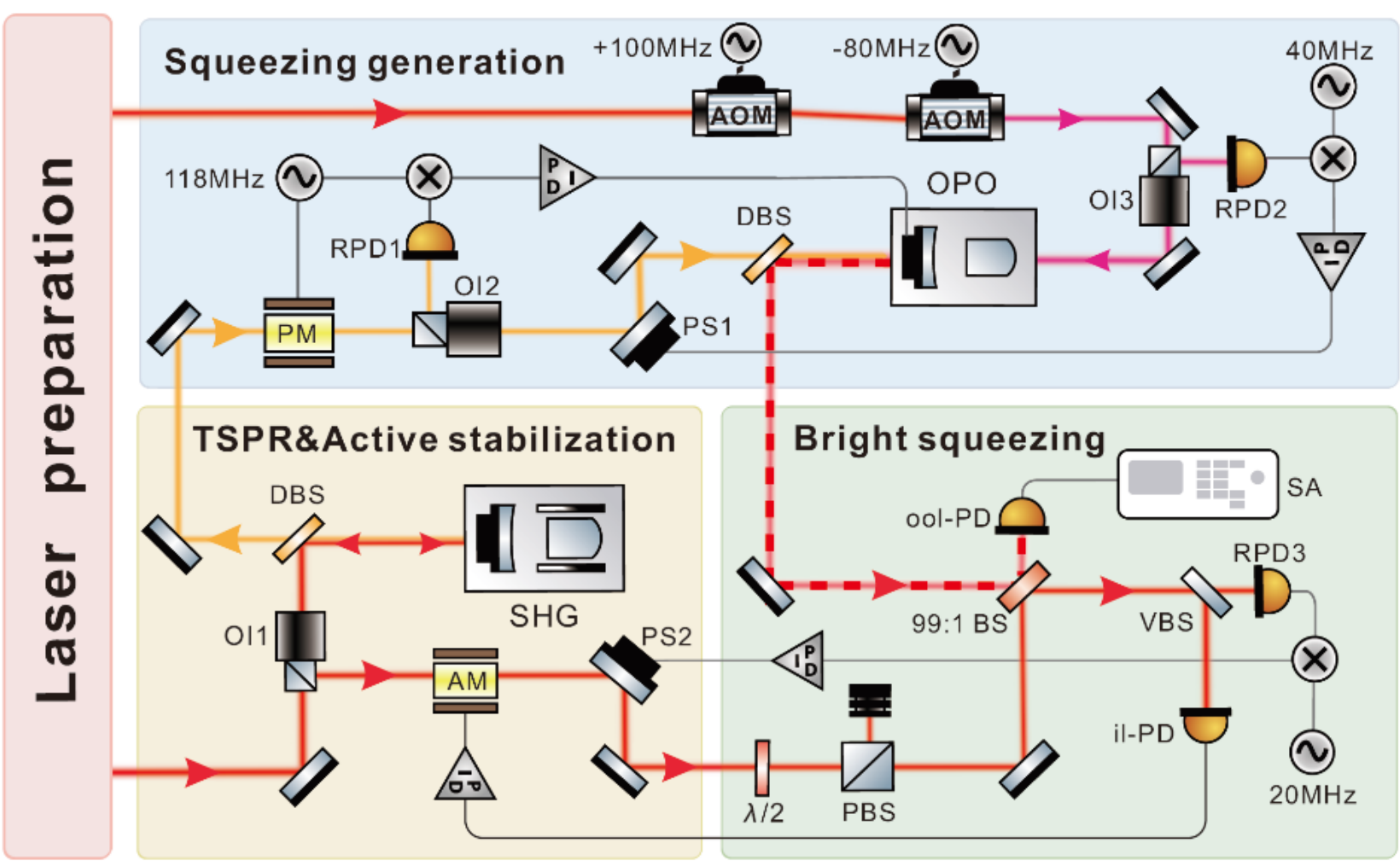

FIG. 3 Experimental setup of the bright squeezed light generation. PM, Phase modulator; AM,

Amplitude modulator; AOM, Acousto-optic modulator; OI, Optical isolator; RPD, Resonant photodetector; PID, Proportional-integral-derivative; SHG, Second harmonic generator; λ/2: Half-wave plate; PBS, Polarization beam splitter; DBS, Dichroic beam splitter; VBS, Variable beam splitter; OPO, Optical parametric oscillator; PS, Phase shifter; il-PD, In-loop photodetector; ool-PD, Out-of-loop photodetector; 99:1 BS, 99:1 beam splitter; SA, Spectrum analyzer.

Fig. 3 illustrates the schematic diagram of bright squeezed light generation. It mainly includes three units: (1) a hybrid technical noise suppression scheme, consists of a passive stabilization technology based on a TSPR technology with a SHG, and a nonclassical active stabilization technology; (2) a squeezed vacuum generator includes an OPO for squeezed vacuum state preparation and a coherent control loop for relative phase locking; (3) a bright squeezed state generation and characterization part. The single frequency laser source used for squeezed light preparation is a 1550 nm continuous-wave (CW) fiber laser (NKT, Koheras BASIK X15) with 1 W output power. Before downstream applications, the laser firstly transmits into a mode cleaner (MC) in laser preparation stage to improve the purity of the laser spatial fundamental mode and polarization, and also filter the laser amplitude and phase noises above the linewidth of the MC[50]. Whereafter, about 500 mW of the laser beam is injected into the SHG with a conversion efficiency of 70%. The generated second harmonic wave of 775 nm acts as the pump source of OPO to generate a squeezed vacuum state. The residual 1550 nm laser reflected from SHG serves as a passively stabilized laser beam with a power of about 110 mW, which is isolated by OI1[49]. Then, it passes through an amplitude modulator (AM) actuator, and is coupled onto a 99:1 beam splitter (99:1 BS) with the squeezed vacuum. The reflected interference field serves as the sensing beam of the in-loop unit, and the transmission one is utilized for out-of-loop applications or amplitude noise characterization. A half-wave plate (λ/2) and a polarization beam splitter (PBS) between the AM and 99:1 BS are combined together to continuously adjust the incident laser power of the 99:1 BS. Before il-PD, a variable beam splitter (VBS) is used for laser attenuation. Its output is injected into the il-PD, which photocurrent is fed back onto the AM to carry out a nonclassical active noise stabilization. The in-loop and out-of-loop PDs are commercial products (Newport model 2053) with saturation power of 10 mW and electronic noise of -170 dB/Hz, whose response frequency is across 1 kHz to 3 MHz. In our experiment, we replace the photodiodes in the PDs with high quantum efficiency ones, which are custom-made by Laser Components GmbH in German (detailed parameters can be found in "Materials and Methods"). A Proportion Integration Differentiation (PID) controller (Vescent model D2-125) with 2 MHz bandwidth is used to realize large gain feedback for amplitude noise suppression.

A squeezed vacuum state is prepared by a sub-threshold degenerate OPO via the phase sensitive parametric down-conversion process. Its squeezing strength is -10.5 dB at the pump power of 16 mW, by budgeting the losses and phase fluctuations of the system. The OPO with a threshold power of 20 mW operates in doubly resonance with both 1550 nm and 775 nm wavelengths, which contains a periodically poled potassium titanyl phosphate (PPKTP) crystal and a planoconcave mirror, and more details can be found in Refs. [51-53]. The 775 nm laser is modulated by a phase modulator (PM) to produce a pair of phase sidebands of ±118 MHz, which enters into the OPO to serve as the pump source and sensing beam for cavity length locking with Pound-Drever-Hall (PDH) technology. A coherent control technique is applied for stabilizing the squeezing angle to produce a stable amplitude squeezed state[54,55]. A 20 MHz frequency-shifted 1550 nm laser beam with 1 mW acts as the phase sensing beam, which is generated by two acousto-optic modulators (AOMs) driven by a

+100 MHz and a -80 MHz sinusoidal signals, respectively. The error signal is demodulated at 40 MHz, and is fed back on the phase shifter (PS1) to lock the relative phase between the pump and squeezing fields to π. While the relative phase between the squeezed vacuum and in-loop sensing beam is locked to zero by the same technique. Here, the pump power of the OPO is reduced to half of its threshold power, if not the larger anti-squeezing noise will lead to instable in the coherent control loop. Therefore, only -8.6 dB squeezed vacuum is utilized in the nonclassical stabilization regime (see Supplementary Note 4). Finally, a bright amplitude squeezed light can be continuously generated at the transmission port of 99:1 BS, which is the out-of-loop beam of our nonclassical stabilization scheme.

## Experimental Results

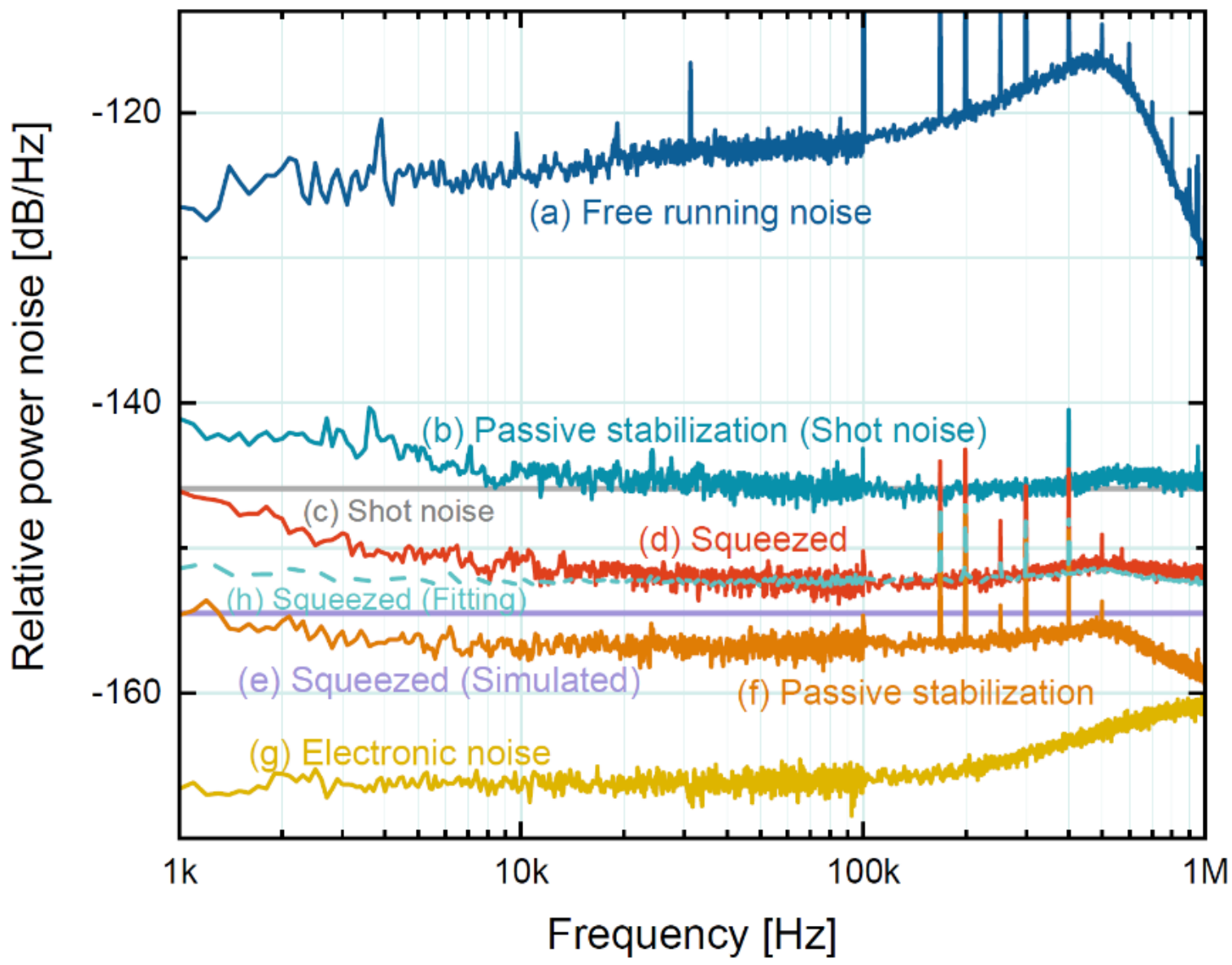

FIG. 4 Measurement results of bright squeezed light at 100 μW with only passive noise stabilization. Trace (a): Measured technical noise of the free-running laser. Trace (b): Measured RSN of the 100 μW detected power of ool-PD, corresponds to the shot noise reference. Trace (c): Theoretical shot noise of a detected power of 100 μW. Trace (d): Measured noise variance of the bright squeezed light, which is 6.5 dB quantum noise reduced among 2 kHz to 1 MHz. Trace (e): Simulated noise variance of the bright squeezed light. Trace (f): Technical noise of the laser under independent passive noise stabilization. Trace (g): Electronic noise of the ool-PD. Trace (h): Uncorrelated sum of the noise sources (e), (f), (g). All measurements are performed with a spectrum analyzer (R&S FSW) at Fourier frequencies from 1 kHz to 100 kHz, and 100 kHz to 1 MHz with resolution bandwidth (RBW) of 100 Hz and video bandwidth (VBW) of 10 Hz.

First of all, we coherently combine a passively stabilized laser beam of 10 mW with the squeezed vacuum of -10.5 dB on the 99:1 BS. As a result, a -6.5 dB bright squeezed light with output power of 100 μW was generated among 1 kHz to 1 MHz frequency range, as show in Fig. 4. Trace (a) depicts the technical noise of the free-running laser at 100 μW. Trace (b) is the shot noise level measured by the ool-PD after passive noise stabilization across the kHz band, compared with the -146 dB/Hz SNL of 100 μW laser (trace (c)). Due to a small cavity detuning of the SHG, the noise within 10 kHz is slightly above SNL. The residual technical noise after passive stabilization (trace (f)) is -157 dB/Hz, corresponding to a detected power of 9.9 mW, which is 11 dB below the SNL (trace (c)). The electronic noise of the ool-PD (trace (g)) is -167 dB/Hz, which is about 20 dB lower than the SNL. The total technical noise of our scheme is the sum of the residual optical technical noise and PD's electronic noise. Apparently, the passive noise stabilization can be applied to generate a bright squeezed light across the kHz to MHz band. But, the measured squeezing strength -6.5 dB deviates from the ideal value predicted by Eq. (1). It attributes to the fact that, except for the technical noise, optical losses and phase fluctuations relating to the squeezing preparation and propagation, are inevitably drawn into the system. The losses introduce vacuum noise, and phase fluctuations project anti-squeezing quadrature into squeezing one. The comprehensive result is reduced the squeezing strength to a lower level (see Supplementary Note 4). The total loss and phase fluctuation[56] are evaluated to $l_{tot} = 0.1$ and the $\theta_{tot} = 21$ mrad, respectively. By simultaneously considering the total technical noise in Table 1, we theoretically simulate the squeezing strength based on Eq. (1), shown as trace (h). The experimental result agrees well with the theoretical one. In fact, the bandwidth of the bright squeezed light is identical with that of the squeezed vacuum, which is limited by the linewidth of the OPO. Here, we only exhibit the quantum noise frequency range up to 1 MHz band.

TABLE. 1 Budget of technical noise normalized to shot noise in passive and multiple stabilizations

| **Source of technical noise (@50kHz)** | Passive (dB) | Passive & active (dB) |
|---|---|---|
| Amplitude noise | -11 | -10 |
| Electronic noise | -21 | -14 |
| Residual in-loop noise | | -30 |
| Total noise | -10.6 | -8.5 |

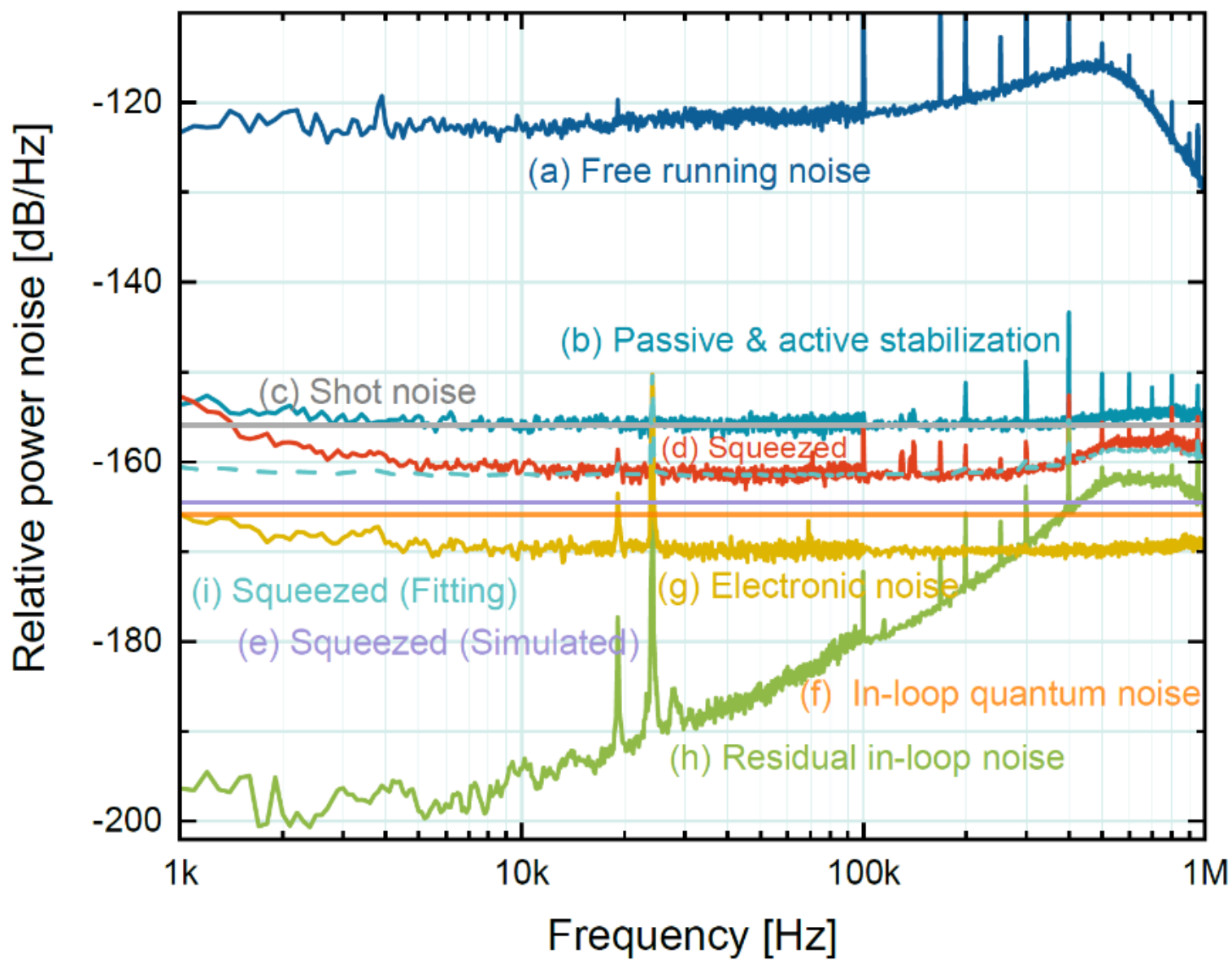

FIG. 5 Measurement results of bright squeezed light at 1 mW with passive and nonclassical active noise stabilizations. Trace (a): Measured technical noise of the free-running laser. Trace (b): Measured RSN of the 1 mW detected power of ool-PD, corresponds to the shot noise reference. Trace (c): Theoretical shot noise at 1 mW. Trace (d): Measured noise variance of bright squeezed light. Trace (e): Simulated noise variance of the squeezed light. Trace (f): The quantum noise coupled through the in-loop vacuum port of the VBS. Trace (g): Sum of the electronic noise of the two-PDs.

Trace (h): Residual technical noise of the in-loop under multiple noise stabilizations. Trace (i): Uncorrelated sum of the noise sources (e), (f), (g), (h). All measurements are performed with a spectrum analyzer (R&S FSW) at Fourier frequencies from 1 kHz to 100 kHz, and 100 kHz to 1 MHz with RBW of 100 Hz and VBW of 10 Hz.

Subsequently, with the hybrid passive and active noise stabilization, we increase the output power of the bright squeezed light to 1 mW. Fig. 5 shows the measured results of the technical and quantum noises across 1 kHz to 1 MHz frequency band. Trace (a) depicts the amplitude noise of the free-running laser at 1 mW. Trace (b) indicates that the stabilized amplitude noise reaches the theoretically predicted SNL (trace (c), -156 dB/Hz at 1 mW). The residual technical noise of the system is reduced to the level of trace (h), which is more than 24 dB below the SNL of the out-of-loop beam within 100 kHz. It is rapidly raised up beyond 100 kHz, due to the frequency dependent feedback gain reduction at higher Fourier frequency. Trace (f) represents the vacuum noise coupled from the empty port of the VBS in the optical path of in-loop unit, which becomes a main influence factor of the squeezing level (see Supplementary Note 2). The electronic noise is measured to be – 170 dB/Hz shown as trace (g). The total technical noise of the out-of-loop beam is the sum of traces (f), (g) and (h), which remains below the SNL. Finally, a -5.5 dB bright amplitude squeezed light (trace (d)) with power of 1 mW is generated among 2 kHz to 1 MHz band. Based on Eqs. (1)-(3) and Eq. (S16) (see Supplementary Note 2), as well as the technical noises in Table 1, we numerically simulate the squeezing level (trace (i)), which agrees well with the experimental one (trace (d)). The noise deterioration below kHz may attribute to three key factors: i) poor performance of the gain and electronic noise of our photodetectors in this frequency range; ii) larger amplitude noise of the laser and electronic noise of the photodetector below kHz induce the SHG locking unstable, iii) noise transferring from the upstream MC.

The results in Fig. 4 infer that the squeezed light generated by single TSPR passive stabilization regime is mainly confined by the technical noise of the laser. Due to the intrinsic loss introduced by the BS and finite noise suppression ability, the passive interference technology with TSPR is only suitable to highly unbalanced BS scheme for low power level squeezing generation (see Supplementary Note 3). But this regime enables to broaden the bandwidth of the squeezed state to MHz frequency band. Further power scaling lowers the $RSN_{OOL}$ approaching to the technical noise floor of the laser as shown in Fig. 1(b), which cuts down the gap between the SNL and technical noise to reduce the squeezing strength, or even completely destroys the squeezed state. In nonclassical hybrid noise stabilization regime, active feedback control can compensate the finite noise suppression ability of the passive one, and enhances the technical noise suppression gain. As a result, the hybrid noise stabilization regime simultaneously possesses the maximum noise suppression ability of the active scheme and broadband noise reduction property of the passive scheme. Hence, the bright squeezed light can be prepared with higher power and squeezing levels in the kHz frequency band shown in Fig. 5. Further improving the optical power, squeezing strength and bandwidth are mainly limited by the detectable power and electronic noise of the il-PD, and bandwidth of the PID controller. The former one sets the lower bound of technical noise, and the later one determines the upper bound of the feedback bandwidth. We can also extend the Fourier frequency of bright squeezing in hundreds of hertz or below by introducing an active feedback control loop to suppress the technical noise in this range[57] before the SHG.

# Discussion

We have demonstrated a novel bright amplitude squeezed light source by integrating a TSPR-based passive stabilization in the nonclassical active feedback scheme. The hybrid dual-stage noise stabilization strategy relaxes the requirements for high gain feedback control loop, and apparently broadens the feedback bandwidth to MHz range. We have also established a universal and complete theoretical model to demonstrate this regime, which can be used to calculate the actual parameters for nonclassical feedback stabilization. Finally, a -5.5 dB bright squeezed light is experimentally prepared with kHz to MHz frequency band and 1 mW output power. To the best of our knowledge, it is the first experimental realization of a milliwatt-order bright squeezed light across the kHz frequency band. To further improve the squeezing or power level for the requirements of different application scenario, we should balance the splitting ratio of the BS, technical and shot noises in the system based on the theoretical models. Our bright squeezed light source can be applied to a wide range of quantum metrology scenarios, such as biological tracking, cavity-mechanics strong coupling, quantum radiation pressure noise suppression, quantum enhanced displacement sensing, and so on.

# Materials and Methods

## Low-phase-noise single frequency fiber laser

A single frequency CW laser NKT (X15) with a linewidth below 100 Hz is used as the laser source. The laser exhibits excellent low phase noise performance, which is benefit for phase sensitivity operation in the bright squeezed light generation. Because of low phase noise enable to minimize the phase noise coupling in the relative phase locking processes, which reduce the phase fluctuation to maintain the squeezing strength transferred from the initial squeezed vacuum state.

## Passive noise stabilization with SHG

The SHG is a monolithic semi-monolithic cavity, which consists of a concave mirror driven by a piezoelectric transducer (PZT) and a periodically poled PPKTP crystal (dimensions: 10 mm × 2 mm × 1 mm). The convex surface of the crystal has a curvature radius of 12 mm coated with high reflectivity (HR) for both 1550 nm and 775 nm wavelengths, while the planar surface is anti-reflection (AR) coated for both wavelengths. The concave mirror has a curvature radius of 30 mm, a transmissivity of 12 ± 1.5% for 1550 nm and high transmission (HT) for 775 nm. The air gap between the two optical elements is 27 mm, and corresponds to a total cavity length of approximately 39 mm. The linewidth and finesse of the SHG are 68 MHz and 48, respectively. The mode-matching efficiency of the 1550 nm laser beam is optimized to ~99.6%. A resonant electro-optic phase modulator (EOPM) drives a pair of ±36 MHz sidebands for PDH technique to stabilize the SHG cavity length. The phase-matching condition is realized by stabilizing the PPKTP crystal's temperature to 36°C with a temperature controller.

During the experiment, a 1550 nm laser beam with a power of 500 mW is injected into the SHG to up-convert a 775 nm laser, which can be used as the pump source of the OPO. Additionally, the SHG also acts as a passive noise stabilization function to suppress the amplitude noise among the kHz band. The physical mechanism relies on the frequency conversion process, which induces a nonlinear relationship between the intracavity circulating power and the reflected fundamental wave. A maximum amplitude noise reduction of 35 dB is achieved as the slope of this nonlinear transfer function approaches to zero corresponding to the conversion efficiency of 70%, at which the noise is reduced to near the SNL among the kHz-MHz frequency range[49]. The power of the passively stabilized laser beam is 110 mW, which is directly applied to the downstream experiments.

## Active noise stabilization with a high gain broadband feedback loop

We employ a broadband PID controller (Vescent, Model D2-125) with 2 MHz bandwidth and a gain adjustable photodetector (Newport, Model 2053, 10 MHz bandwidth) to construct an 80 dB flat gain feedback control in kHz frequency range. We replace the photodiode of the detector with a high quantum efficiency one (>99%) to minimize the loss in squeezed state detection. It has a 100 μm diameter of active area, and is 20° AR coated with 1550 nm wavelength for s-polarization. The detector also has a low electronic noise of -170 dB/Hz. An electro-optic amplitude modulator (EOAM) (Thorlabs: EO-AM-NR-C3) acts as an actuator for laser technical noise stabilization. The amplitude noise of its output laser beam is sensed by the il-PD installed on the reflected port of the 99:1 BS. To avoid excess noise coupling, no external amplifier is used to drive the EOAM. With these optimizations, a high gain feedback control loop is realized with 1 kHz to 1 MHz frequency bandwidth.

**Acknowledgements**

The project is sponsored by National Natural Science Foundation of China (NSFC) (Grants No. 62225504, No. U22A6003, No. 62027821, No. 62375162), National Key Research and Development Program of China (No. 2024YFF0726401).

**Author contributions**

R.-X.L., Y.-J.W. and Y.-H.Z. design the experiment. R.-X.L., N.-J.J. and B.-N.A. carry out the experiment with assistance from Y.-J.W. and L.-R.C. B.-N.A. and J.-Y.L. help collect the data. R.-X.L., Y.-J.W. and Y.-H.Z. analyze the data and write the paper with input from all other authors. The project is supervised by Y.-H.Z. All authors discuss the experimental procedures and results.

**Data availability**

The authors declare that all data supporting the findings of this study can be found within the paper. Additional data supporting the findings of this study are available from the corresponding author (Y.J.W and Y.H.Z.) upon reasonable request.

**Conflict of interest**

The authors declare no competing interests.

# Supplementary Information for:

# Bright squeezed light in the kilohertz frequency band

Ruixin Li[1], Bingnan An[1], Nanjing Jiao[1], Junyang Liu[1], Lirong Chen[1,2], Yajun Wang[1,2,*], and Yaohui Zheng[1,2,**]

[1]*State Key Laboratory of Quantum Optics Technologies and Devices, Institute of Opto-Electronics, Shanxi University, Taiyuan 030006, China*

[2]*Collaborative Innovation Center of Extreme Optics, Shanxi University, Taiyuan 030006, China.*

**YJWangsxu@sxu.edu.cn*

***yhzheng@sxu.edu.cn*

This document provides supplementary materials for "Bright squeezed light in the kilohertz frequency band". It contains four parts: (1) Theory for bright squeezed light generation via passive interference; (2) Theory for bright squeezed light generation via nonclassical active feedback; (3) Comparation of passive interference and active feedback regimes; (4) Loss and phase fluctuation budget for hybrid nonclassical stabilization.

## 1. Theory for bright squeezed light generation via passive interference

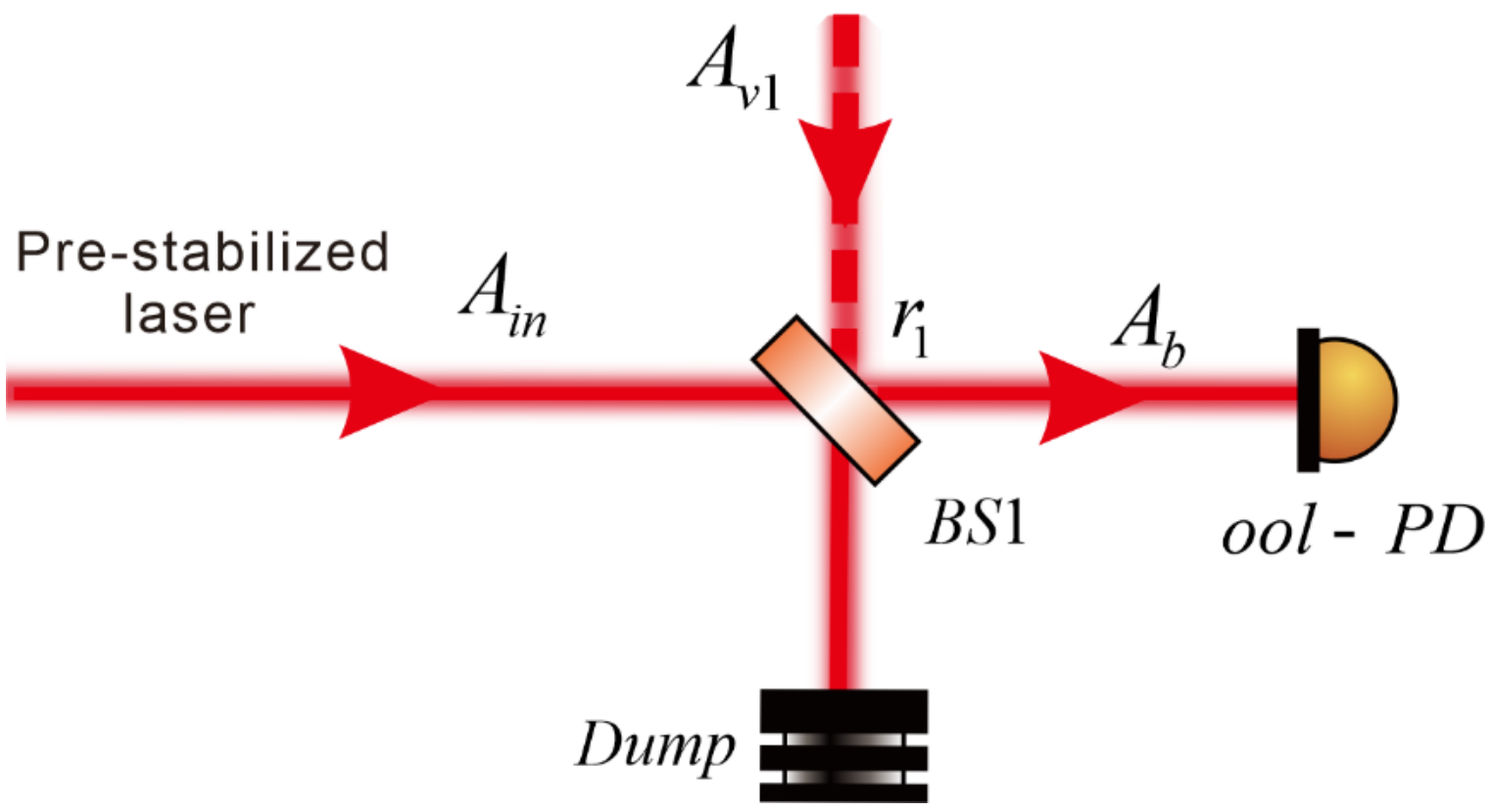

Fig. S1 Schematic diagram of bright squeezed light generation via passive interference. A simplified illustration on how a pre-stabilized laser is interfered with a squeezed vacuum at a beam splitter (BS1) to generate a bright amplitude-squeezed light, which supports the derivation of Eqs. (S1)–(S6).

Figure S1 illustrates the theoretical model for bright squeezed light generation using a pre-stabilized[1,2] field $A_{in}$, which is interfered with a squeezed vacuum at a beam splitter (BS1). The amplitude quadrature fluctuation of the output field is given by:

$$\delta A_b = -\sqrt{r_1}\delta A_{v1} + \sqrt{1-r_1}\delta A_{in}, \quad (S1)$$

here, $r_1$ is the power reflectivity of BS1, $\delta A_{in}$ is the amplitude noise of the input field, $\delta A_{v1}$ is the squeezed vacuum fluctuation.

The variance of the output field $\delta A_b$ becomes:

$$V_b = r_1 V_{v1} + (1-r_1)V_{in}\,. \quad (S2)$$

The shot and technical noises make up the input noise $V_{in} = V_{sn} + V_{tn}$, and substitute into Eq. (S2):

$$V_b = r_1 V_{v1} + (1-r_1)(V_{sn} + V_{tn}). \quad (S3)$$

We multiply the relative shot noise of the output field, $RSN_{OOL} = 2h\nu/P_{OOL}$, on both sides of Eq. (S3)[3], where $h$ is Planck constant, $\nu$ is the laser frequency, and $P_{OOL}$ is the optical power of the output field, and it yields:

$$RSN_{OOL}V_b = RSN_{OOL}r_1 V_{v1} + (1-r_1)RSN_{OOL} + g(f)\cdot TN_F, \quad (S4)$$

where $g(f) = V_{output}/V_{input}$ is a factor of the passive noise suppression by three-step photon recycling (TSPR) stabilization[1], where $V_{input}$ and $V_{output}$ are the noise variances before and after stabilization, respectively. $TN_F$ is the relative technical noise of the free-running laser.

After rearranging Eq. (S4), we obtain the total noise variance of the bright squeezed field at the output port of BS1:

$$V_b = r_1 V_{v1} + (1-r_1) + \frac{TN_{OOL-P}}{RSN_{OOL}}, \quad (S5)$$

where $TN_{OOL-P} = g(f)\cdot TN_F$ represents the residual technical noise after passive stabilization. As the power reflectivity of the BS1 approaches to unity ($r_1 \approx 1$), the variance of the generated bright squeezed light can be approximated as:

$$V_b \cong V_{v1} + \frac{TN_{OOL-P}}{RSN_{OOL}}. \quad (S6)$$

**2. Theory for bright squeezed light generation via nonclassical active feedback**

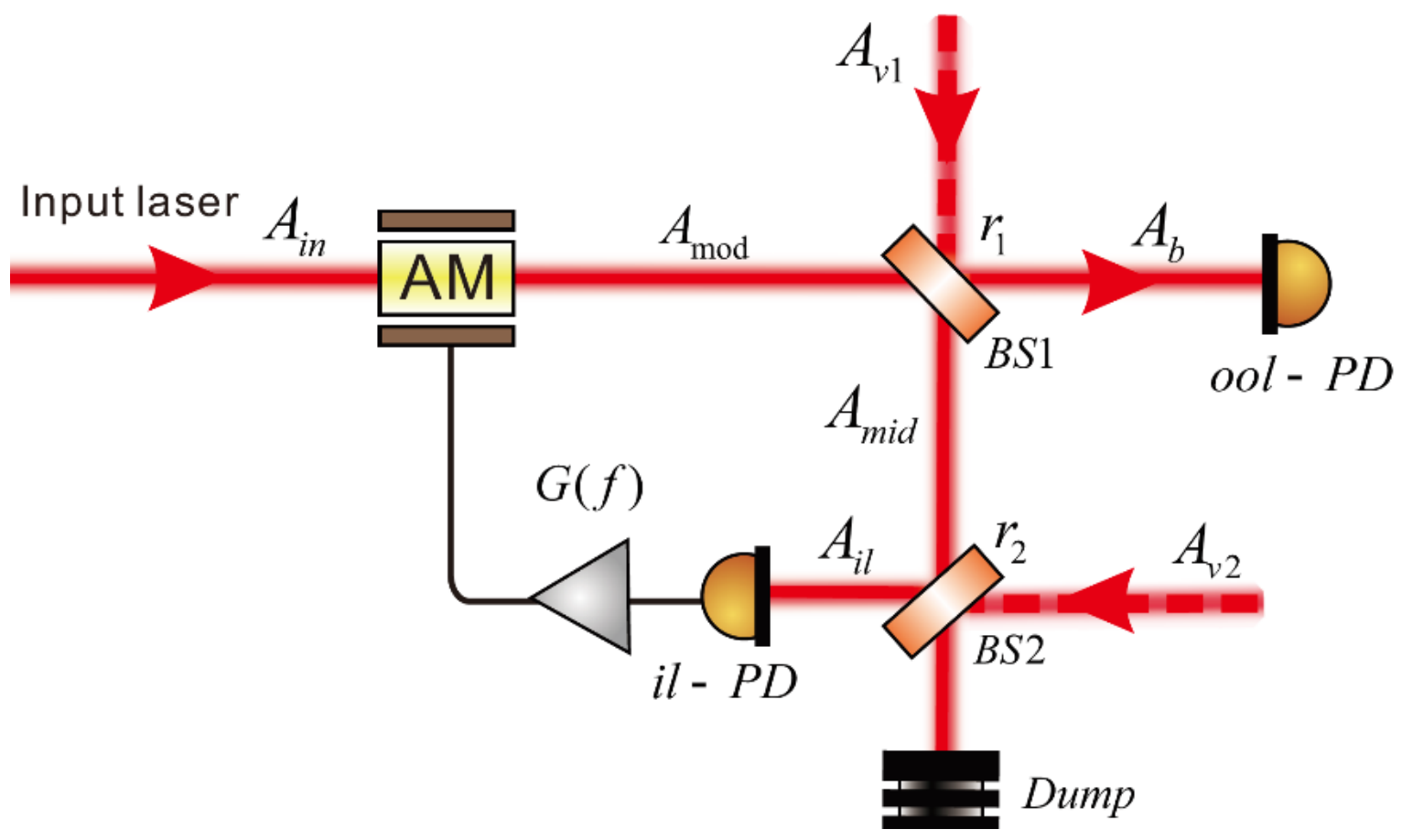

Fig. S2 Schematic diagram of bright squeezed light generation via nonclassical active feedback. Detailed representation of the feedback loop incorporating squeezed vacuum injection via BS1, power attenuation via BS2, and closed-loop feedback. This model serves as the basis for the derivation of Eq. (S7) – (S17).

Figure S2 shows the theoretical model for the nonclassical active feedback scheme. The input laser beam $A_{in}$ enters into a nonclassical active feedback control loop, and interferes with a squeezed vacuum at BS1. Then, after being attenuated by BS2, it is detected by in-loop photodetector (il-PD). The output current of il-PD is feedback on to the upstream amplitude modulator (AM) for actively feedbacking the quadrature noise. The relation between these fields can be described by[4-6]:

$$\delta A_{mod} = \delta A_{in} + G(f)\delta A_{ilPD}, \tag{S7}$$

$$\delta A_{mid} = \sqrt{r_1}\delta A_{mod} + \sqrt{1-r_1}\delta A_{v1}, \tag{S8}$$

$$\delta A_{il} = \sqrt{r_2}\delta A_{mid} + \sqrt{1-r_2}\delta A_{v2}\ , \tag{S9}$$

$$\delta A_{ilPD} = \delta A_{il} + \delta e_{ilPD}\ , \tag{S10}$$

$$\delta A_b = \sqrt{1-r_1}\delta A_{mod} - \sqrt{r_1}\delta A_{v1}\ , \tag{S11}$$

here, $G(f)$ is the gain of the nonclassical active feedback system, which represents the frequency-dependent noise suppression ability within the feedback bandwidth; $\delta A_{ilPD}$ is the amplitude noise readout by the il-PD, and $\delta A_{mod}$ is that of the output field of AM; $r_1$ and $r_2$ are the power reflectivities of the first and second BSs, respectively; $\delta A_{v1}$ and $\delta A_{v2}$ denote the squeezed vacuum and vacuum noise entering through the empty ports of the two BSs, respectively; $\delta A_{mid}$ is the noise of the optical field between BS1 and BS2; $\delta A_{il}$ represents the amplitude noise entering into the il-PD; $\delta e_{ilPD}$ is the electronic noise of the il-PD.

By substituting Eqs. (S7)–(S10) into (S11), we obtain the output field fluctuation $\delta A_b$ as:

$$\delta A_b \cong \frac{\sqrt{1-r_1}\delta A_{in} + \delta A_{v1} G(f)\sqrt{r_2}}{(1-G(f)\sqrt{r_1 r_2})} +$$

$$\frac{G(f)\sqrt{1-r_1}\sqrt{1-r_2}\delta A_{v2} + \sqrt{1-r_1}G(f)\delta e_{ilPD}}{(1-G(f)\sqrt{r_1 r_2})}. \quad (S12)$$

With the feedback gain $G(f) \gg 1$, the output noise variance becomes:

$$V_b \cong \frac{(1-r_1)V_{in}}{(1-G(f)\sqrt{r_1 r_2})^2} + \frac{V_{v1}}{r_1} + \frac{(1-r_1)(1-r_2)V_{v2}}{r_1 r_2} + \frac{(1-r_1)V_e}{r_1 r_2}. \quad (S13)$$

Multiplying the relative shot noise of the output field $RSN_{OOL}$ on both sides of Eq. (S13), we obtain the relative noise form as[3]:

$$RSN_{OOL}V_b = \frac{RSN_{in}V_{in}}{(1-G(f)\sqrt{r_1 r_2})^2} + RSN_{OOL}\frac{V_{v1}}{r_1} + RSN_{IL}(1-r_2)V_{v2} + RSN_{IL}V_e, (S14)$$

here, $RSN_{in}$ and $RSN_{IL}$ are the relative shot noise of the input laser beam and the il-PD, respectively. The noise of the input field includes the shot and technical noises, $V_{in} = V_{sn} + V_{tn}$, and also has a form of $RSN_{in}V_{in} = RSN_{in} + g(f)\cdot TN_F$. Then, the noise variance of the generated bright squeezed light can be expressed as:

$$V_b = \frac{V_{v1}}{r_1} + \frac{\frac{RSN_{in} + g(f)\cdot TN_F}{(1-G(f)\sqrt{r_1 r_2})^2} + RSN_{IL}(1-r_2)V_{v2} + RSN_{IL}V_e}{RSN_{OOL}}. \quad (S15)$$

In our scheme, the input laser is passively pre-stabilized, after which a nonclassical feedback control is carried out. In general, $V_{tn} \gg V_{sn}$, the technical noise of the output field after the hybrid passive and active stabilizations can be defined as:

$$TN_{OOL-P\&A} \cong \frac{g(f)\cdot TN_F}{(1-G(f)\sqrt{r_1 r_2})^2} + RSN_{IL}(1-r_2)V_{v2} + RSN_{IL}V_e. \quad (S16)$$

The first term of Eq. (S16) represents the residual technical noise of the out-of-loop unit, which can be completely removed with an infinitely large feedback gain $G(f)$. The second term is the quantum noise relating to the power reflectivity $r_2$ of BS2 in the in-loop unit, which comes from the vacuum noise coupled from BS2 and relative shot noise of the detected laser beam in the in-loop unit. The last term is the electronic noise of the feedback control loop.

When the entire optical power is detected by the il-PD, i.e., $r_2 \approx 1$. Eq. (S16) is simplified to Eq. (3) in the main text. Under nonclassical active stabilization, the noise variance of the output bright squeezed light can be expressed as:

$$V_b = \frac{V_{v1}}{r_1} + \frac{TN_{OOL-P\&A}}{RSN_{OOL}}. \quad (S17)$$

### 3. Comparation of passive interference and active feedback regimes

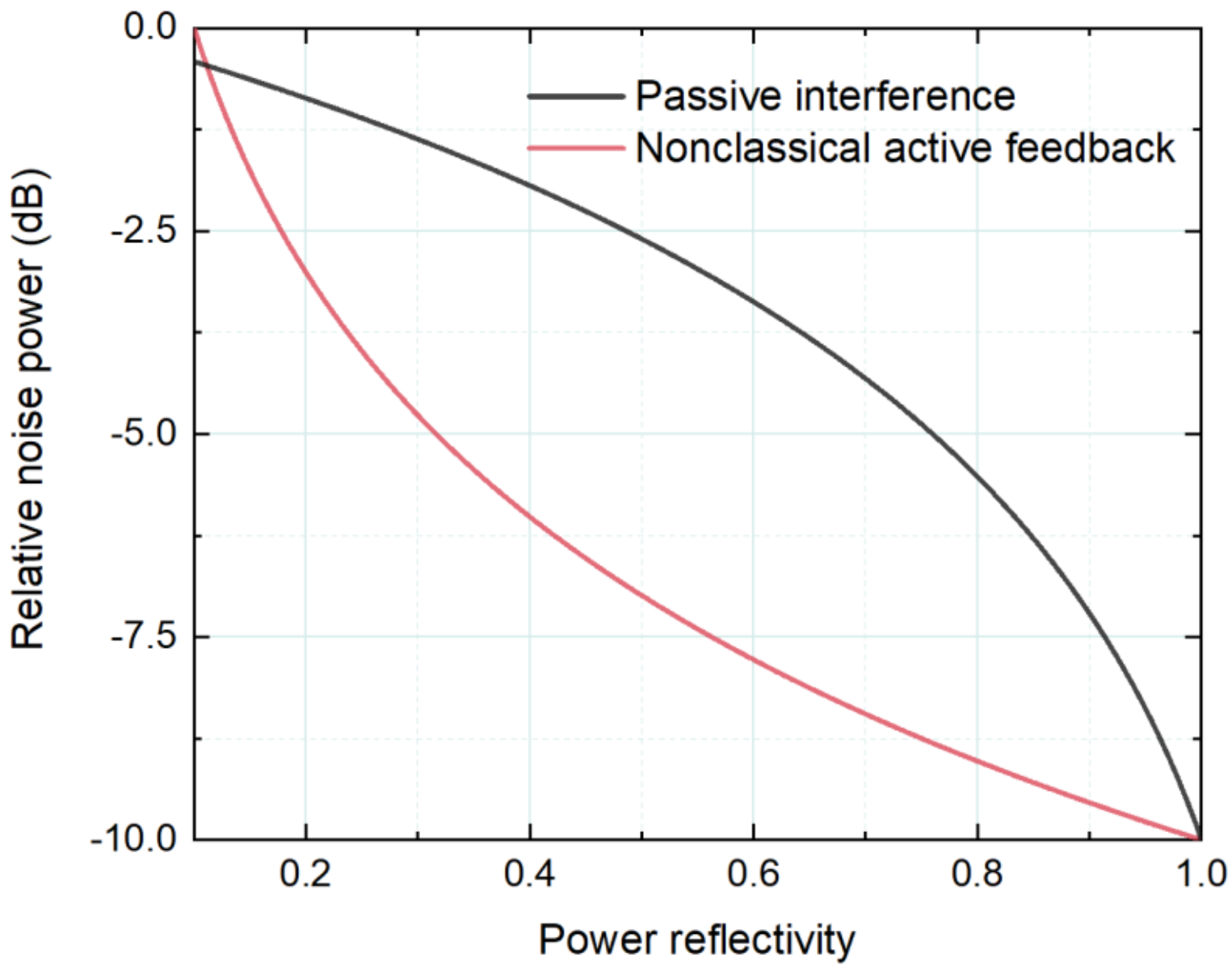

Fig. S3 Noise power of the output field of BS1 for passive interference and nonclassical active feedback schemes with different power reflectivity of BS1.

TABLE. S1 Examples of passive interference and nonclassical active feedback regimes for bright squeezing generation at different splitting ratio ($r_1$: $t_1$) of BS1.

| BS1 ratio ($r_1$: $t_1$) | Squeezing strength (dB) | | Squeezing power (mW) |
|---|---|---|---|
| | Passive interference | Nonclassical active feedback | |
| 99:1 | -9.63 | -9.96 | 1 |
| 90:10 | -7.21 | -9.54 | 10 |
| 50:50 | -2.60 | -6.99 | 50 |
| 20:80 | -0.87 | -3.01 | 80 |
| 10:90 | -0.41 | 0 | 90 |

We compare the passive interference and nonclassical active feedback regimes as a function of the power reflectivity of BS1 for bright squeezed light generation. We assume that the technical noise of a 100 mW input laser beam is completely removed, and a -10 dB squeezed vacuum is coupled on BS1 (see Figs. S1 and S2). Based on Eqs. (S5) and (S17), the squeezing strengths of the two schemes are calculated as shown in Fig. S3, and Table S1 summarizes the squeezing strength and power properties for them at different splitting ratio ($r_1$: $t_1$) of BS1.

Apparently, as the power reflectivity decreases, more optical power can be extracted to scale the power of the bright squeezed light. However, squeezing strength is reduced gradually due to increased vacuum noise coupling. From Fig. S3 and Table S1, we can infer that the nonclassical feedback scheme always owns stronger squeezing strength than passive interference one at different splitting ratio, except $r_1 < 0.11$. Even with a highly imbalanced BS, e.g., 99% reflectivity, the nonclassical active feedback regime still presents a 0.33 dB noise suppression enhancement than the passive interference one. We can enlarge the difference between the two regimes by changing the splitting ratio to a balanced BS, which is also benefited to increase the power level of the nonclassical light. Therefore, in order to balance the squeezing and power levels in different application scenario, we should change the splitting ratio of the BS.

## 4. Loss and phase fluctuation budget for hybrid nonclassical stabilization

Table S2 presents the optical loss and phase fluctuation budget in our experiment. The total optical loss, including escape efficiency of optical parametric oscillator (OPO), interference visibility, quantum efficiency of photodiode and propagation loss, is approximately 10%, which introduces vacuum noise in the squeezing quadrature, and limits the observable squeezing. While the total phase fluctuation is about 21 mrad, which degrades the measured squeezing by mixing the anti-squeezing quadrature into the squeezing one.

TABLE. S2 Budget of the optical loss and phase fluctuation relating to the squeezing preparation

| **Source of optical loss** | Relevant loss (%) |
|---|---|
| OPO escape efficiency | 3.0±0.3 |
| Efficiency of interference | 2.8±0.2 |
| Quantum efficiency of photodiodes | 1.0±0.2 |
| Propagation efficiency | 3.2±0.5 |
| Total efficiency | 10±0.8 |
| **Source of phase fluctuation** | Value (mrad) |
| OPO | 2±0.1 |
| Relative phase between squeezed and frequency-shifted light | 8±0.2 |
| Relative phase of squeezed and local field | 11±0.5 |
| Total phase fluctuation | 21±0.8 |

By considering the total optical loss $l_{tot}$ and phase fluctuation $\theta_{tot}$, the variances

$V_{a/s}$ of anti-squeezing and squeezing quadratures can be expressed as[7,8]

$$V_{a/s} = \left[1 \pm \frac{4(1-l_{tot})\sqrt{p/p_{th}}}{\left(1 \mp \sqrt{p/p_{th}}\right)^2 + 4(f/k)^2}\right] cos^2\,\theta_{tot} + \left[1 \mp \frac{4(1-l_{tot})\sqrt{p/p_{th}}}{\left(1 \pm \sqrt{p/p_{th}}\right)^2 + 4(f/k)^2}\right] sin^2\,\theta_{tot}\,, \qquad (S18)$$

where $p$, and $p_{th}$ are the pump and threshold powers of the OPO, respectively; $f$ and $k$ denote the Fourier frequency and cavity linewidth of the OPO, respectively. Based on Eq. (S18), $l_{tot} = 10\%$, $\theta_{tot} = 21$ mrad, $p = 10$ mW, $p_{th} = 20$ mW and $k = 110.4$ MHz, the observed squeezing strength of -8.6 dB agrees well with the theoretical value.